\documentclass[12pt,a4paper]{article}
\usepackage[utf8]{inputenc}
\usepackage{amsmath}
\usepackage{amsfonts}
\usepackage{amssymb}
\usepackage{graphicx}
\usepackage[left=2cm,right=2cm,top=2cm,bottom=2cm]{geometry}
\begin{document}
\begin{center}
\large{\textbf{Encoding by DNA Relations and Randomization Through Chaotic Sequences for Image Encryption}}\\
\normalsize{Chiranjoy Chattopadhyay\footnote{chiranoy@gmail.com}, Bikramjit Sarkar\footnote{sarkar.bikramjit@gmail.com}, Debaprasad Mukherjee\footnote{mdebaprasad@gmail.com, Corresponding Author}\\
Depts. of Computer Science and Engineering \& Information Technology,\\ Dr. B. C. Roy Engineering College, Durgapur, West Bengal, India}
\end{center}
\textbf{Abstract: }Researchers in the field of DNA based chaotic cryptography have recently proposed a set of novel and efficient image encryption algorithms. In this paper, we present a comprehensive summary of those techniques, which are available in the literature. The discussion given in this paper is grouped into three main areas. At first, we give a brief sketch of the backbone architecture and the theoretical foundation of this field, based on which all the algorithms were proposed. Next, we briefly discuss the set of image encryption algorithms based on this architecture and categorized them as either encryption or cryptanalyzing techniques. Finally, we present the different evaluation metrics used to quantitatively measure the performance of such algorithms. We also discuss the characteristic differences among these algorithms. We further highlight the potential advances that are needed to improvise the present state-of-the-art image encryption technique using DNA computing and chaos theory. The primary objective of this survey is to provide researchers in the field of DNA computing and chaos theory based image encryption a comprehensive summary of the progress achieved so far and to facilitate them to identify a few challenging future research areas.
\section{Introduction}
\label{sec:intro}
At present, digital image is one of the most important modes of sharing information across the globe. Therefore, it is important to secure them from illegal interception, tampering and destruction. There are mission critical systems like military image databases, medical imaging system, video conferencing, television, as well as personal albums in social networking sites, which require highly secured systems to store, retrieve and transmit digital images. The primary goal of any digital image encryption research is to fulfill this need of designing a highly secured system to prevent unauthorized use of digital images.

The work on cryptographic research has its root in the seminal work on permutation-diffusion pattern by Shannon \cite{shannon1949communication}. Since then, researchers have worked towards designing robust image encryption algorithms. Due to some intrinsic features of images such as bulk data capacity and high redundancy, traditional encryption schemes appear not to be idle for images and thus new directions of information security are being sought to protect the data. Recently, Bio-computing methods are applied as a potential alternative for image encryption. Researchers have drawn parallels between properties of chaotic systems and that of cryptographic systems. For example, the mixing property of a chaotic system is similar to the cryptographic property, which says that a small perturbation to the local area generates huge change in the entire space. DNA computing is another field of study that uses the recombinant DNA techniques for performing computation. DNA cryptography is a newly evolved field of study and at the same time very promising. It uses DNA for information storage and transmission.
  
Since $1990$s, due to the close relationship between chaos theory and cryptography, researchers have moved their attention towards chaos-based cryptosystems. However, various studies have proved that encryption algorithms based only on chaotic maps, have lower key space and not so robust against different types of statistical and image processing attacks. On the other hand, DNA cryptography is a new and promising field in cryptography that has appeared with the development of DNA computing and exploits DNA to carry information with the assistance of molecular methods. Last five years have a seen a novel trend among the researchers to combine the goodness of both these field of study, i.e. chaos theory and DNA computing. Researchers have proposed algorithms based on the combined concepts of chaos theory and DNA computing in such a way that resulting cryptographic systems are having the good qualities of both the fields.

\subsection{Objective of the paper}
The objective of this paper is to provide an overview of the major techniques on DNA based chaotic encryption of digital images. In this paper, we will discuss about the backbone architecture and the theoretical foundation based on which all the algorithms were proposed. Especially, we will give a detailed discussion on the individual techniques that were proposed in different scenarios and provide a comprehensive collection of the recent publications in this field. We will also present the different performance evaluation metrics used in this field and their significance. Furthermore, we will point out the pros and cons of these algorithms and highlight the potential advances that are needed to improvise the present state-of-the-art. Image encryption techniques, which are solely based on chaos theory or DNA encoding scheme are not in the scope of the paper.

\subsection{Organization of the paper} The paper is organized in the following way. We present the general overview of the DNA based chaotic encryption techniques in Sec. \ref{sec:overview}. A brief description of the different encryption and cryptanalyzing algorithms are given in Sec. \ref{sec:desc}. We postpone the discussion on the different performance evaluation metric till Sec. \ref{sec:metric}. After that, we highlight the pros and cons of the different algorithms in Sec. \ref{sec:discuss}. The paper concludes with a summary and the potential future scope of works in Sec. \ref{sec:summ}.
\section{General Overview}
\label{sec:overview}
\begin{figure}
\begin{center}
\includegraphics[scale=.75]{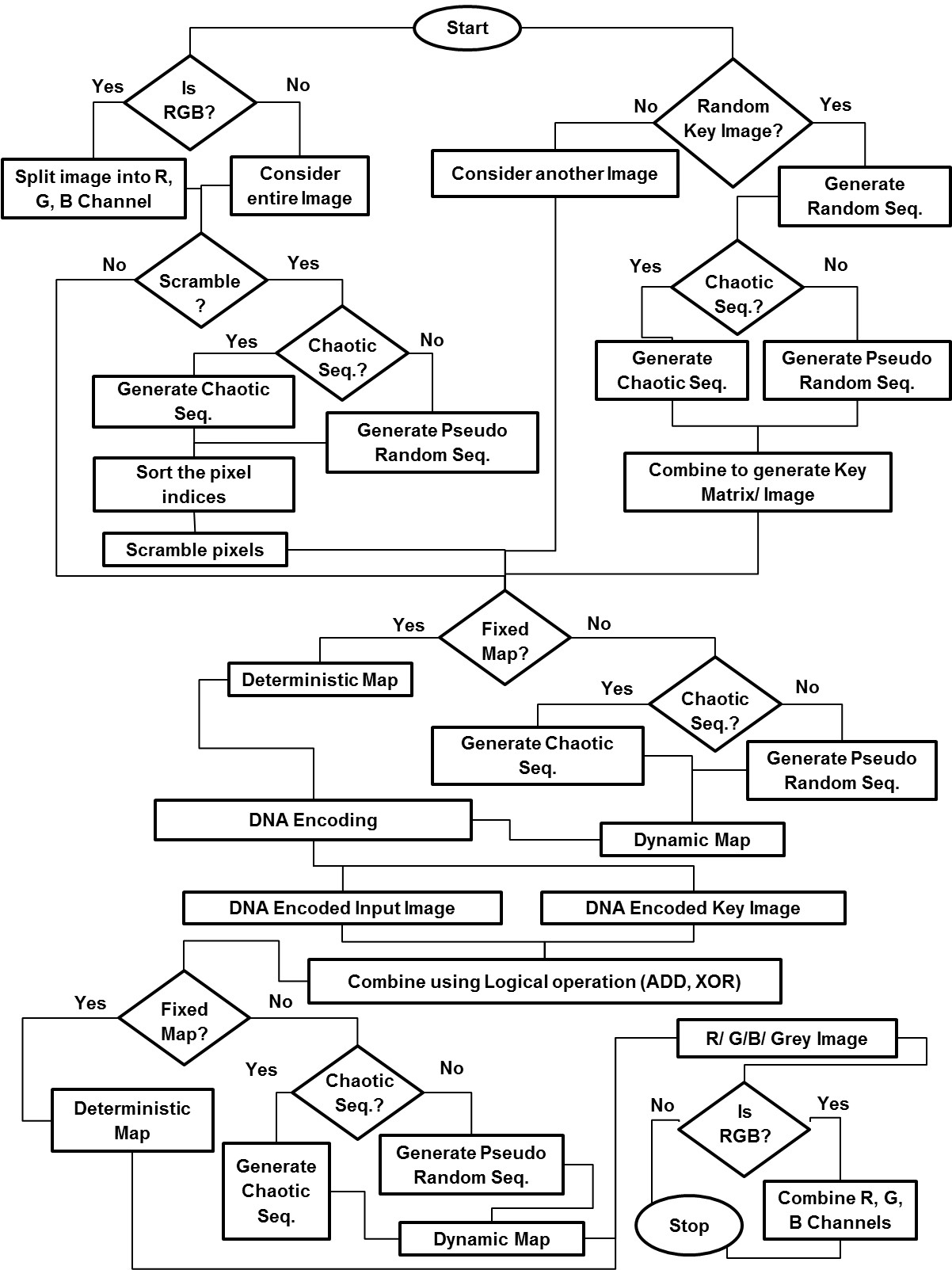}
\end{center}
\caption{A flow chart of different approaches of DNA based chaotic cryptography techniques.}
\label{fig:flowchart}
\end{figure}

In the literature, there exists several DNA based chaotic cryptographic technique (detail discussion is given in Sec. \ref{sec:desc}) . In Fig. \ref{fig:flowchart}, we present a flow chart that summarizes the methods discussed in all those papers in one place. This flow chart is the backbone architecture of all the algorithms. A brief discussion of this flow chart is given next.

\begin{figure}[t]
\begin{center}
\includegraphics[scale=.55]{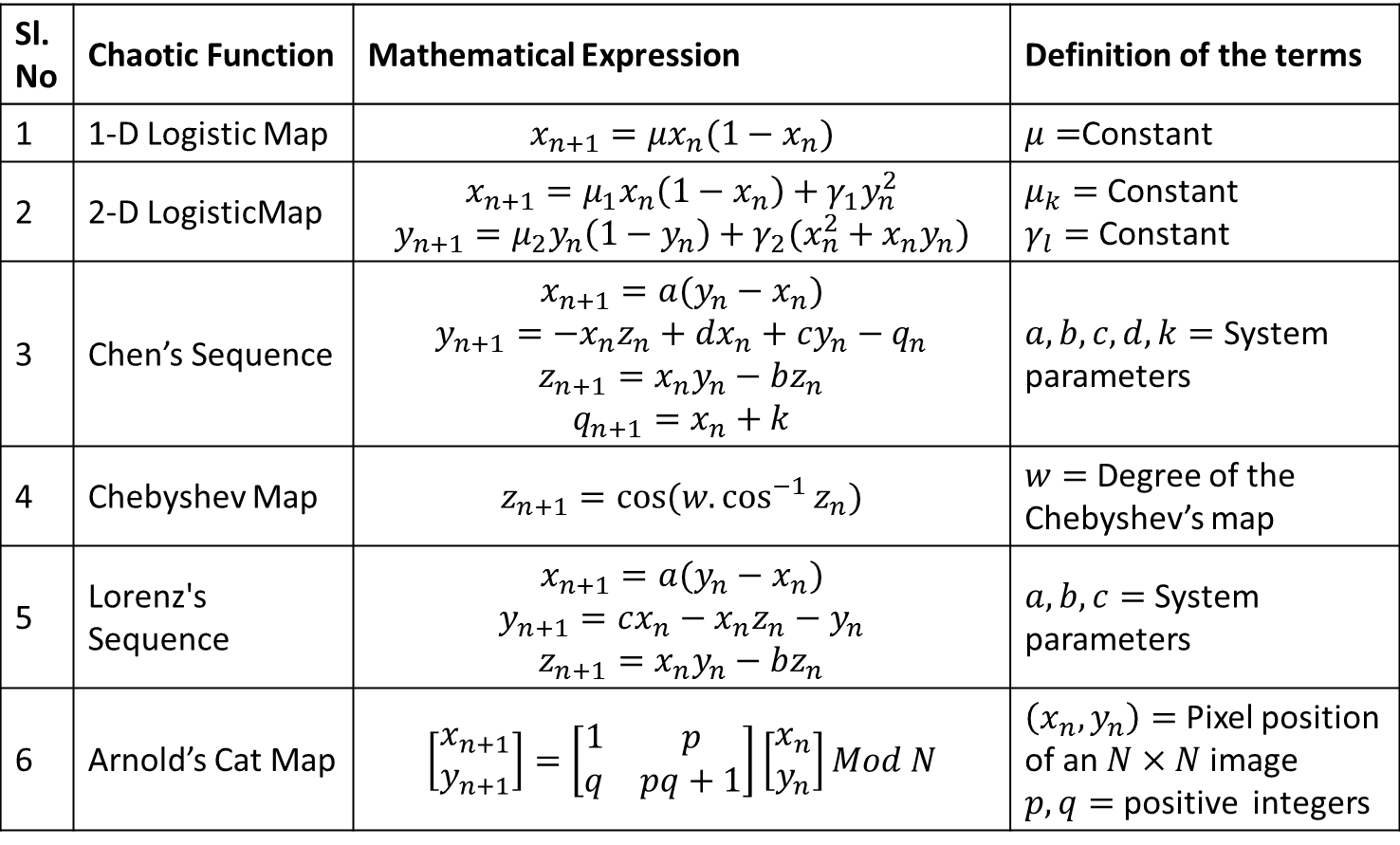}
\end{center}
\caption{Different chaos maps used in the literature to generate chaotic sequences. In all the mathematical expressions above left hand side of the expression denotes the next term in the sequence.$n$ = number of iterations.}
\label{fig:chaos_eq}
\end{figure}

There are two parallel paths in the algorithm (as shown by the two elbow joints from the ''start'' symbol in Fig. \ref{fig:flowchart}). They are: DNA encryption of the (i) original input image , and (ii) secret key image. Later, these two images are combined to obtain the final encrypted image. An input image is of two types, (i) color (RGB), and (ii) gray-scale. A color image is composed of three channels, i.e. R, G and B, whereas a gray-scale image has only a single channel. Henceforth, an image channel will be referred as image matrix. For color images, the encryption algorithm is applied to the individual matrices and then the resultant image is obtained by combining these matrices. However, for gray-scale images the algorithms work only on the single matrix. On the other hand, the secret key image could be a original image (generally gray-scale) or it could be a synthetic image generated using chaos map. Figure \ref{fig:chaos_eq} summarizes the different chaos maps and there corresponding mathematical expressions to generate the chaotic sequences.

In general, an image encryption technique is divided into two stages: (i) confusion and (ii) diffusion. During the confusion stage, position of the image pixels are scrambled (changed) to make the original image unrecognizable. However, it has to be ensured that the original image can be obtained by performing the reverse operations. Image pixels are scrambled by at first generating a pseudo-random sequence either using chaos maps or by random number generator. This sequence is used to change the pixel locations and thereby introduce confusion. The next stage is to encrypt the scrambled image, i.e. diffusion.

\begin{figure}
\begin{center}
\includegraphics[scale=.55]{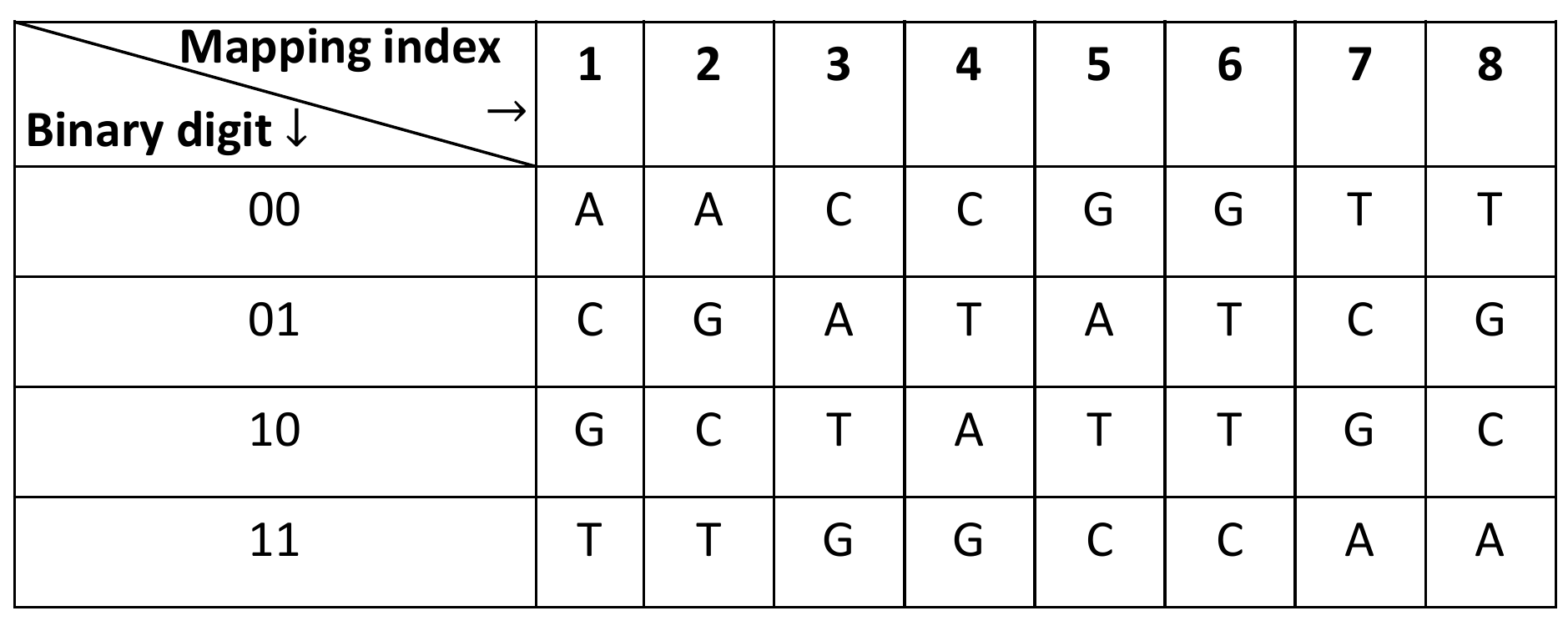}
\end{center}
\caption{Eight different permissible binary to DNA map because of the complementary relationship between the bases A,G, T and C.}
\label{fig:dna_map}
\end{figure}

 The diffusion stage essentially alters the individual values stored in the pixel locations and thereby breaks the strong correlation among the neighboring pixels. The encryption algorithm is applied to the input image matrix and the secret key matrix. Each pixel value of a image matrix is DNA encoded. The basic element of DNA (nucleotide) is divided into four chemical components (bases), they are: (i) Adenine (A), (ii) Guanine (G), (iii) Thymine (T), and (iv) Cytosine (C). Each pixel value (an $8$ bit number) is divided into $4$ binary numbers of $2$ bits. These numbers are mapped to their corresponding DNA sequence. Because of the complementary relationship among the four bases, out of all possible combination $(4!=24)$, only $8$ different combination can be used (see Fig. \ref{fig:dna_map} for details). A fixed mapping uses one out of these eight mappings for encryption and reverse mapping for decryption. However, a dynamic mapping uses different maps at different level to increase the randomness in the encryption process and thereby makes it more robust. 
 
\begin{figure}[b]
\begin{center}
\includegraphics[scale=.55]{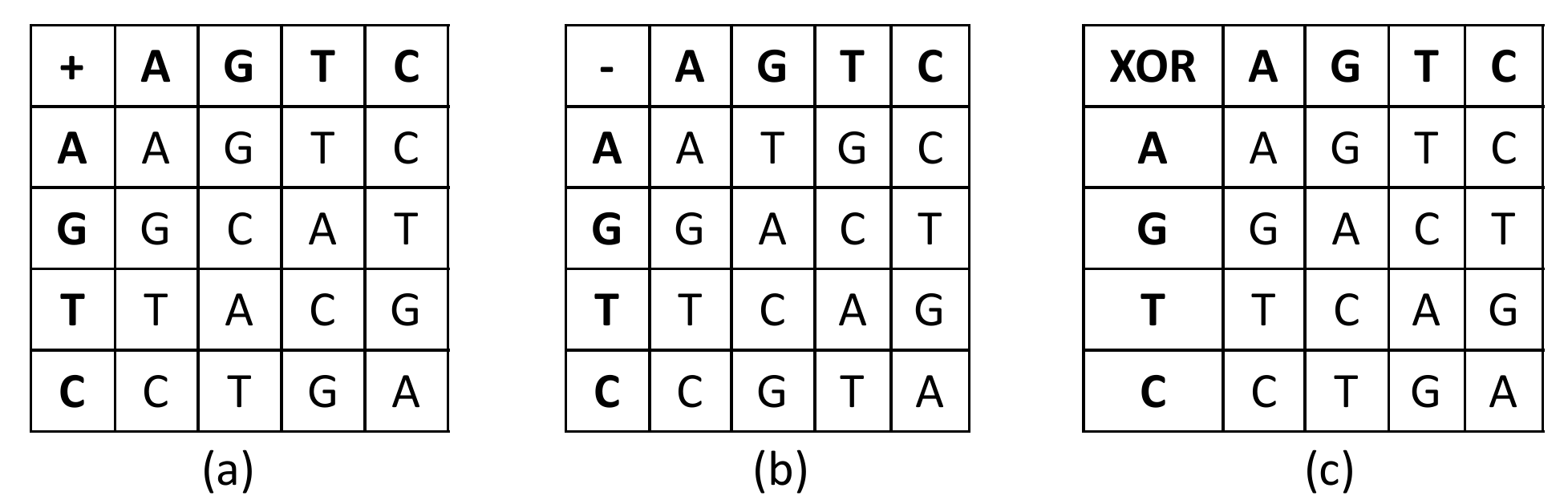}
\end{center}
\caption{One type of logical operation on DNA sequences, (a) addition, (b) subtraction, and (c) XOR. An entry in the $(i,j)^{th}$ location signifies the result of the corresponding logical operation between the bases at the location $i$ and $j$.}
\label{fig:logop}
\end{figure}
Once the input and the secret key images are DNA encoded, they are combined together using logical operations defined on the DNA strings. After DNA encoding, every pixel in the resultant matrix is defined by a DNA sequence of length $8$. In the literature, logical addition, and XOR operations are used to combine the secret key and input matrix during encryption process. However, logical subtraction and XOR operations are used during the decryption process. Figure \ref{fig:logop} gives the truth tables of such logical operations on DNA bases. After the combination the resultant DNA string is decoded back to binary and then to decimal digits. These decimal values represent the value at the pixel locations of the final encrypted image. For RGB images, the results of encryption on individual image matrices are combined to obtain the final encrypted image. During decryption, a reverse process is applied to get the input image. Following section gives detail descriptions of the major algorithms, which are based on DNA based chaotic encryption techniques on digital images.
\section{Detailed Description of Algorithms}
\label{sec:desc}
\begin{figure}[t]
\begin{center}
\includegraphics[scale=.7]{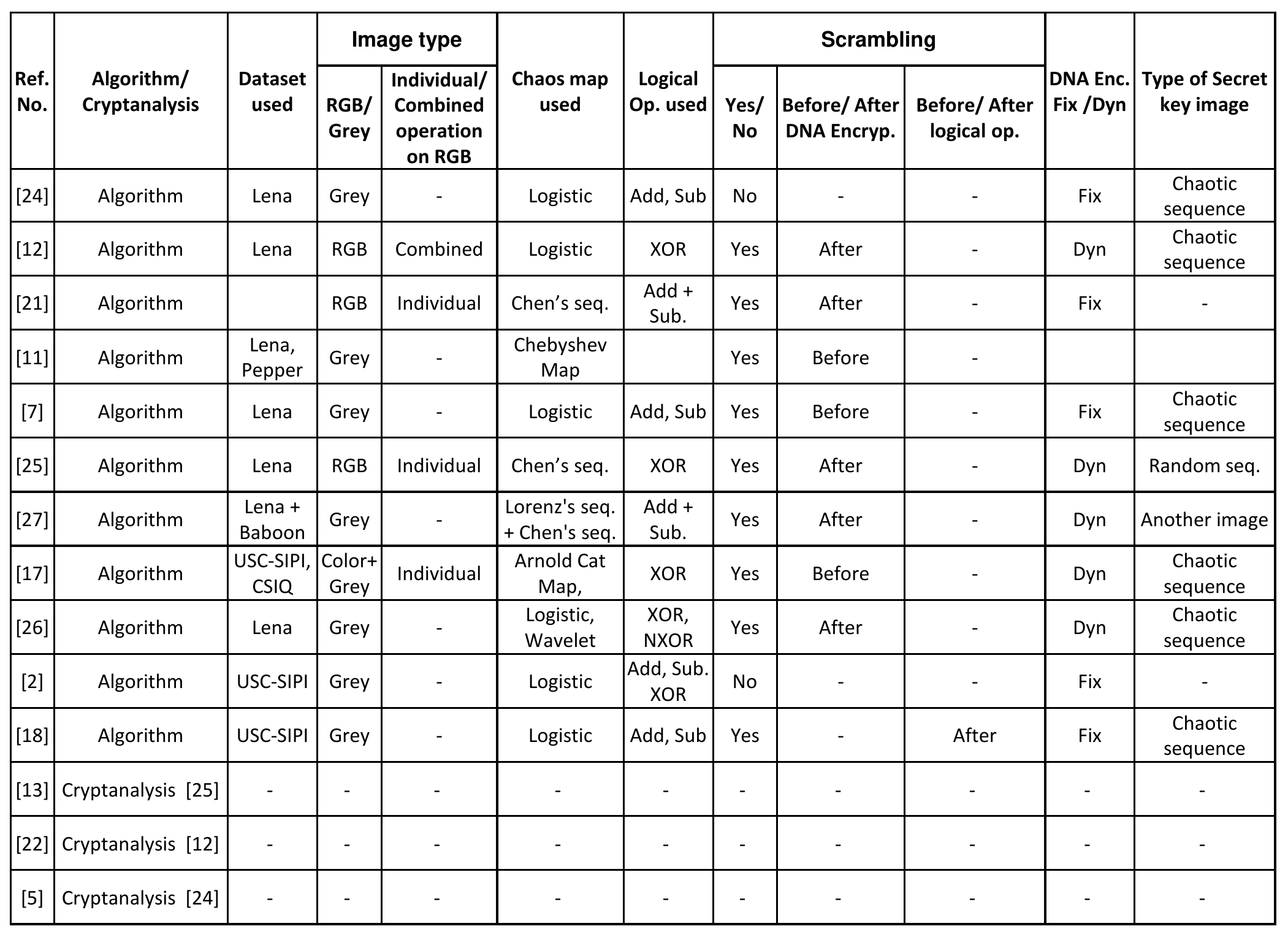}
\end{center}
\caption{A comparison of the different techniques used for digital image cryptography using DNA computing and chaos theory.}
\label{fig:table}
\end{figure}

The field of DNA based image encryption in combination with chaos theory has recently evolved. However, there are works published in the literature which are explicitly based on either chaos theory \cite{gao2008new,xu2012improved,zhou2008parallel,teng2012bit,mazloom2009color,zhang2014image,liu2010color} or DNA encryption \cite{jain2013adaptive,leier2000cryptography,roy2011improved,zhou2010image,zhang2012novel,gehani2004dna}. All the works published in the literature can be grouped into two categories. They are: (i) proposal of a new algorithm, and (ii) cryptanalysis of an existing algorithm. In this paper, we give details of both the categories. Figure \ref{fig:table} highlights the key features of the these algorithms. Each row in this table signifies one algorithm, while the columns denote individual features. The features were chosen based on the general algorithmic flow of these algorithms (see Sec. \ref{sec:overview}). In Fig.\ref{fig:table}, the algorithms are chronologically ordered and also divided into two groups (new algorithms (category (i)) are given in row $2-11$, and cryptanalyzing methods are given in row $13-15$. The references given in the second column of row $13-15$ signifies the cryptographic algorithms, which these papers are analyzing. Entries from column $3$ to $10$ of row $13-15$ are left blank because these features are missing in them. Detail description of the algorithms are discussed in the following sub-sections.
\subsection{Proposal of A New Algorithm}
\label{sub:crynew}
Algorithms in this category have proposed a new technique of image encryption, as well as validation and comparative study. 

The first attempt to combine DNA computing and chaos theory in a unified framework was made by Zhang et al. \cite{zhang2010image}. A new image encryption scheme was proposed in \cite{zhang2010image} using the concept of DNA sequence addition operation and chaos. The algorithm starts with a DNA sequence matrix, which is obtained by encoding the input image. The matrix was then divided into equal blocks of pixels and DNA sequence addition operation was applied to add these blocks. After that, a DNA sequence complement operation was applied to the resultant matrix with the help of two Logistic maps. At the end, the DNA sequence matrix was decoded to generate the encrypted image. Zhang et al. \cite{zhang2010image} have reported the experimental results and security analysis on the well known Lena image (gray-scale). The algorithm was shown to attain high encryption accuracy, as well as immune to exhaustive attack, statistical attack and differential attack.\\
An RGB image encryption algorithm was proposed by Liu et al. \cite{liu2012rgb}. At the beginning of the algorithm, the RGB image was divided into three channels, i.e. R, G, and B. Then, the individual R, G, and B components were transformed into DNA code. Pixel locations of the resultant DNA sequences matrices were scrambled using a Logistic chaos function to disturb the correlation between the pixels in spatial domain. DNA addition operation was carried out between the DNA sequence matrices and then resultant matrices were complemented with DNA sequences with the help of another chaotic sequences. After that, the resultant image matrices were decoded, and scrambled using a Logistic chaos map. Finally, the R, G, B components were combined to obtain the encrypted RGB images. The algorithm proposed by Liu et al. effectively eliminates the correlation between the pixels of the RGB image in the spatial domain, and resistant to different types of attacks.\\
Wei et al. \cite{wei2012novel} have proposed a new color image encryption technique. Their algorithm was based on DNA sequence operations. At first, the R, G, and B channels of an RGB image are DNA encoded using a fixed binary to DNA mapping. After that, pixel locations of the elements of the DNA encoded image components were scrambled using Chen\textsc{\char13}s hyper-chaotic maps. Next, these blocks were added with the help of DNA addition operation and Chen\textsc{\char13}s hyper-chaotic maps. Finally, the DNA sequence matrices were decoded and the individual channels were recombined. Authors have performed simulations and security analysis of their proposed encryption algorithm. Results have demonstrated the goodness of the algorithm, as well as the robustness to different kind of attacks.\\
A novel confusion and diffusion method was proposed by Liu et al. \cite{liu2012image} for image encryption. Their algorithm works only for the gray-scale images. At first, a proposed piece-wise linear chaotic map (PWLCM) was used to scramble the rows and columns respectively. Once the image pixels were encoded the complementary rules were applied to transform each nucleotide into their base pair with the help of Chebyshev chaotic maps. MD 5 hash of the plain image was used  to generate the initial values and parameters of the chaotic maps. Simulation results were shown on gray-scale images to demonstrate high encryption quality and robustness against common attacks.\\
Jain and Rajpal \cite{jain2015robust} proposed an image encryption technique using chaotic maps and DNA encoding techniques. At the beginning, the input image was DNA encoded and $1D$ chaotic map was used to generate a mask. DNA addition technique was used to add this mask with the DNA encoded input image. A complement matrix was produced with the help of a pair of $1D$ chaotic maps and applied on the resultant matrix. Finally, $2D$ chaotic map was used to permute the resultant matrix and DNA decoding technique was applied to obtain the cipher image. The authors had shown that their proposed technique is totally invertible and resistant to different kinds of cryptanalyzing attacks.\\
Zhang et al. \cite{zhang2013novel_fusion} proposed a new image encryption algorithm based on DNA sequence operation and hyper-chaotic system. The algorithm is divided into three stages. Stage 1: two DNA sequences matrices were obtained by encoding the original input image and the secret key image. Stage 2: Chen\textsc{\char13}s hyper-chaotic map was used to generate the chaotic sequences. These sequences were used to scramble the elements of the DNA sequence matrix. Stage 3: XOR operation was used to combine the original an the key image. The combined DNA sequence matrix was encoded to generate the encrypted image. Zhang et al. have reported experimental results and detailed analysis to show that their algorithm not only has good encryption effect, but also has the ability of resisting exhaustive attack and statistical attack.\\
Zhang and Wei \cite{zhang2013novel} proposed a novel couple images encryption algorithm based on DNA sub-sequence operation and chaotic system. This algorithm \cite{zhang2013novel} exploits the idea of DNA sub-sequence operation (such as elongation operation, truncation operation, and deletion operation). And then, do the DNA addition operation under the Chen\textsc{\char13}s  Hyper-chaotic map in this image cipher. The simulation experimental results and security analysis show that our algorithm not only has good encryption effect, but also has the ability of resisting exhaustive attack and statistical attack.\\
A chaos based symmetric key encryption technique for RGB color images was proposed by Som et al. \cite{som2013colour}, with the help of DNA coding and a Chaos based Pseudo-random Binary Number Generator (PRBNG). Firstly, generalized Arnold Cat Map  was used to scramble the input image locations and pixel values. The scrambled image pixel are converted to DNA codes and again reconverted to integers. The chaos based PRBNG was used to generate binary sequences to determine the DNA coding rule. An $1D$ Logistic map was used to produce the cipher image and diffused with the DNA encoded input image by performing exclusive OR operation defined on the DNA sequences. The authors have performed both qualitative and quantitative analysis of their proposed algorithm. Experiments were performed on two benchmark image datasets \cite{weberusc,larson2010most} and compared with existing methods to show the robustness of their proposed encryption algorithm.\\
Zhang et al. \cite{zhang2014improved} have proposed an image encryption technique that combines DNA encoding and chaotic systems. Their algorithm uses chaotic system to shuffle the pixel locations and pixel values. After that DNA encoding technique was applied on the resultant image according to the Quaternary code rules, which is governed by the Quaternary chaotic sequences. At the end, the encrypted image was generated by applying DNA decoding technique. Authors have performed both theoretical analysis and experimental evaluations to show the robustness of their proposed algorithm.\\ 
A novel image encryption algorithm was proposed by Enayatifar et al. \cite{enayatifar2014chaos} by combining a hybrid model of DNA masking, a genetic algorithm (GA) and a logistic map. DNA and logistic map functions were used to generate a set of initial DNA masks. The optimum mask for encryption was determined by applying a GA technique. The significant advantage of this approach is that it improves the the quality of DNA masks to obtain the best mask that is compatible with plain images. Authors have used add, subtract and exclusive or logical operations to combine a pair of DNA encrypted images. Similar to Som et al. \cite{som2013colour}, experiments were performed on standard Their proposed technique demonstrate excellent encryption quality as well as high degree of resistant to various typical
attacks.\\
Very recently, Soni and Acharya \cite{soni2012novel} have proposed a hybrid approach of chaos and DNA encoding methods for image encryption. For the two basic steps of an image encryption technique, i.e. confusion (permutation) and diffusion authors have used Chaos sequence and DNA encoding techniques, respectively. 1D logistic map was used to generate index based chaotic sequence for permutation. Then DNA encoding rule was applied to obtain a DNA sequence matrix by encoding the permuted image and index based chaotic sequence. The DNA matrices were combined using DNA addition operation and decoded using DNA decoding rule to obtain the encrypted  image. Two salient features of this algorithm are: (i) the process by which real valued chaotic logistic sequences are converted to integer sequence and (ii) the formation of the encoded DNA key matrix. Experimental results revealed that the proposed algorithm ensures high security, unaffected by statistical attacks. The proposed algorithm has a greater key space to operate with and also very sensitive to the key value.\\
The following sub-section describes the different cryptanalyzing algorithms that measure the effectiveness, and correctness of an already proposed image encryption algorithm.
\subsection{Cryptanalysis of an Existing Algorithm}
\label{sub:cryanal}
The study of the techniques used to break existing cryptosystems is called cryptanalysis \cite{el2013image}. In addition to mathematical analysis of cryptographic algorithms, these algorithms also studies the side-channel attacks that do not target weaknesses in the cryptographic algorithms themselves, but instead exploit weaknesses in their implementation.

Liu et al. \cite{liu2014cryptanalyzing} performed cryptanalysis on the novel image encryption algorithm proposed by Zhang et al. \cite{zhang2013novel_fusion}. Zhang et al. \cite{zhang2013novel_fusion} claimed that their scheme can be broken with $4mn/3 +1$ chosen plain-images and the corresponding cipher-images, where $m$ is the number of rows  and $n$ is the number of columns of the input image. Liu et al. re-evaluates the security of the encryption scheme proposed in  \cite{zhang2013novel_fusion}. Their analysis proves that the algorithm proposed in \cite{zhang2013novel_fusion} can be broken with less than $\lceil log_2(4mn)/2 \rceil +1$ chosen plain-images. To support their claim, Liu et al. have performed detailed theoretical and experimental analysis.\\
Liu et al. \cite{liu2012rgb} have shown that four pairs of chosen plain-images and the corresponding cipher-image is required to break their proposed encryption algorithm. However, Xie et al. \cite{xie2014breaking} have shown that this claim is not true. In their work \cite{xie2014breaking}, Xie et al. have performed rigorous theoretical analysis and experimental evaluations to reveal the shortcomings in the algorithm proposed by Liu et al. The cryptanalysis results revealed the following security issues in the encryption algorithm. They are: (a) only one pair of known plain image is enough to reconstruct the equivalent secret key of the encryption algorithm, (2) encryption results are not sensitive with respect to the changes of the plain-images/secret key.

Hermassi et al. \cite{hermassi2014security} proposed to cryptanalyse the encryption algorithm proposed by Zhang et al. \cite{zhang2010image}, which combines a DNA addition and a chaotic map to encrypt a gray scale image. The primary contribution of Hermassi et al. were: (i) to demonstrate that the algorithm \cite{hermassi2014security} is non-invertible. If an encryption algorithm is not invertible then it is not possible for the receiver to decrypt the cipher image, even with the help of the secret key, (ii) to describe a chosen plain text attack on the invertible encryption block.\\

\section{Evaluation Metrics used}
\label{sec:metric}
The primary characteristic of a good encryption algorithm is to make changes in the input image in such a manner that the difference between the pixel values of the original and the encrypted images maximizes. Moreover, the final encrypted image must not reveal any of the features of the original image. Visual inspection is one way of measuring the difference between an original and encrypted image. However, it is not enough to quantitatively measure the differences an thus evaluation metrics are necessary.

\begin{figure}[t]
\begin{center}
\includegraphics[scale=.5]{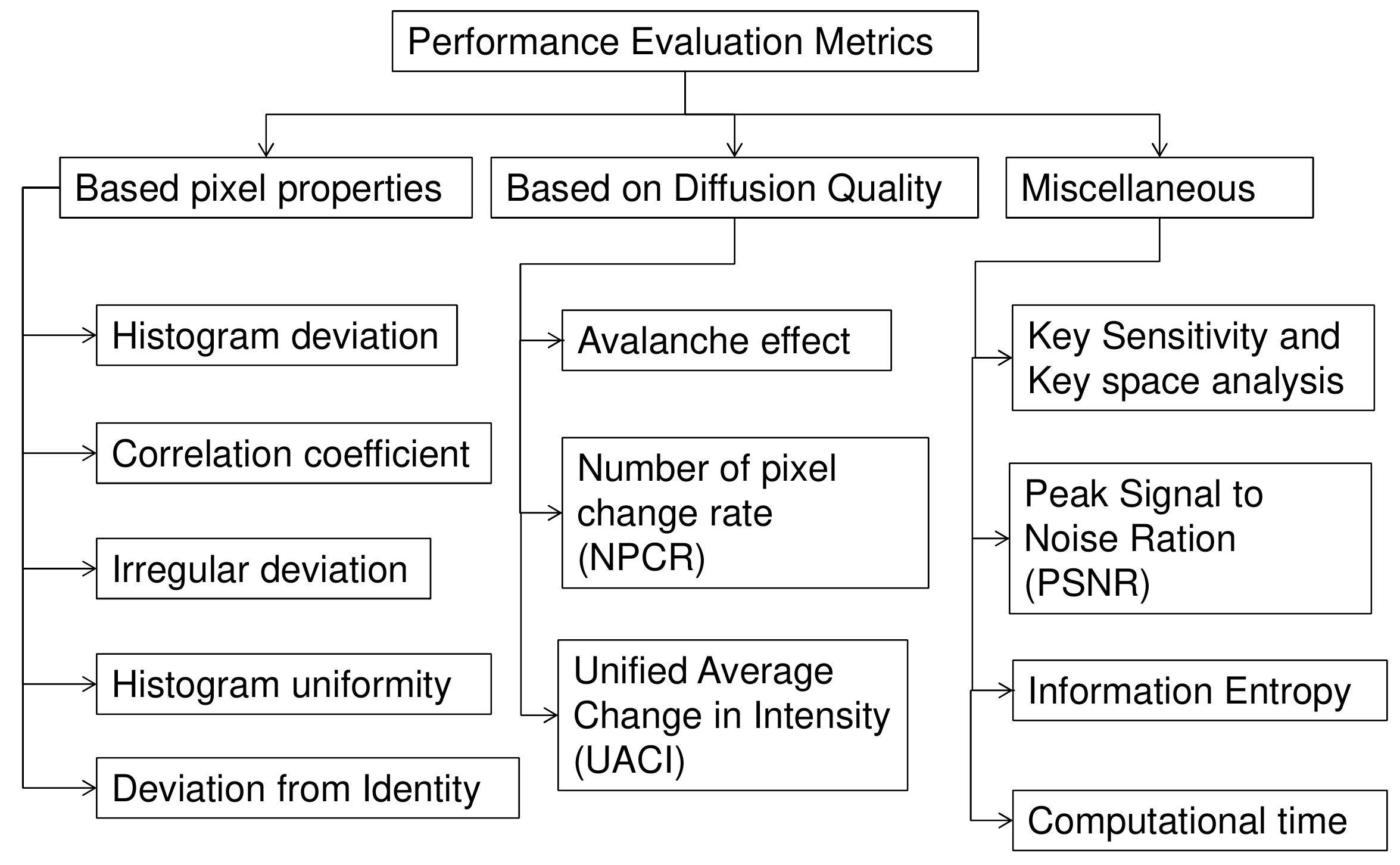}
\end{center}
\caption{Different categories of metrics used to measure the performance of any DNA based chaotic encryption techniques.}
\label{fig:metric}
\end{figure}

\begin{figure}[!b]
\begin{center}
\includegraphics[scale=.48]{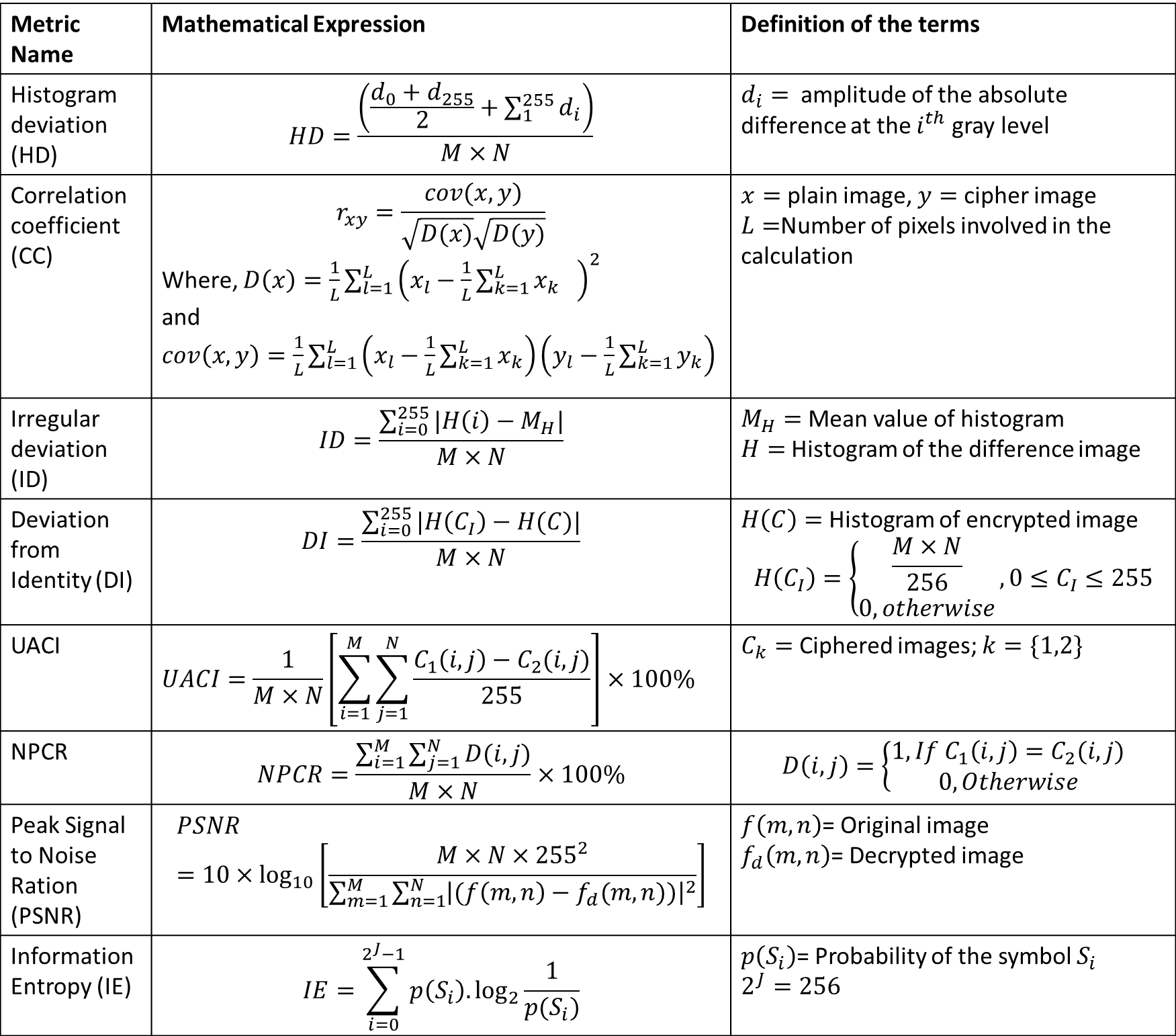}
\end{center}
\caption{Different categories of metrics used to measure the performance of any DNA based chaotic encryption techniques. In all the equations $M =$ image height and $N =$ image width.}
\label{fig:metric_eq}
\end{figure}

Figure \ref{fig:metric} depicts the different evaluation metrics available in the literature to measure the performance of an image encryption algorithm using a combination of DNA computing and chaos theory. The metrics can be grouped into three sets, based on: (i) pixel properties,(ii) diffusion quality, and (iii) miscellaneous. In the following subsections we present the fundamental ideas behind each of these metrics. Details of the mathematical formulas used to compute values of the different evaluation metrics is summarized in a table for ease of understanding (see Fig. \ref{fig:metric_eq}). Each row of the table given in Fig. \ref{fig:metric_eq} corresponds to one metric. The second column shows the mathematical expression,while the third column gives the definition of the terms used in those expressions.
\subsection{Metric Based on Pixel Properties}
\label{subsec:pixel}
The set of metrics in this group evaluates the ability of the encryption algorithm to substitute the original image with an uncorrelated encrypted image. The different metrics under this category are:
\subsubsection{Histogram Deviation (HD)}
\label{subsub:hd}
The histogram deviation (HD) measures the deviation between the original and the encrypted images. Higher the value of HD reflects better encryption accuracy. Even though this is a fairly reliable measure of the difference between the original and the encrypted image, it is not sufficient to use only this as a measure of encryption quality because it depends on the difference between the histograms of the input and the cipher image.
\subsubsection{Correlation coefficient}
\label{subsub:cc}
A convenient metric to measure the quality of any image encryption algorithm is the correlation coefficient between pixels at the same location in the original and the cipher images. It is desirable that the correlation coefficient value between the original and the encrypted image should be close to zero, indicating that the pair of images is uncorrelated. This ensures that the proposed algorithm is guarded against the pixel correlation statistical attack.
\subsubsection{Irregular deviation (ID)}
\label{subsub:id}
The irregular deviation (ID) measures the quality of encryption in terms of how much the deviation caused by encryption (on the encrypted image) is irregular. A difference image is calculated by calculating the absolute difference between the input and encrypted image. Histogram analysis is performed on this difference image. Lower value of ID indicates higher encryption accuracy.
\subsubsection{Histogram Uniformity (HU)}
\label{subsub:hu}
Suppose $H_I$ be the histogram of the input image and $H_{I'}$ be the histogram of the encrypted image. Then, $H_{I'}$ should have the following characteristics: (i) $H_{I'}$ should be totally different from $H_{I}$ , and (ii) $H_{I'}$ must have a uniform distribution.
\subsubsection{Deviation from Identity (DI)}
\label{subsub:di}
The histogram of an ideally encrypted image should have a uniform distribution of all the gray levels. The DI metric measures the deviation of the histogram of the encrypted image from the histogram of an image, which is ideally encrypted. Lower value of DI reflects better encryption quality.
\subsection{Metric based on diffusion quality}
\label{sub:diff}
The set of metrics in this group evaluates the efficiency of the diffusion mechanism. Different metrics in this category are:
\subsubsection{Avalanche effect (AE)}
\label{subsub:ae}
Suppose, a single bit is altered in the original input image $I$. The resultant image is denoted by $I'$. Now, both $I$ and $I'$ are encrypted to give $I_E$ and $I'_E$. The AE metric measures the percentage of different bits between $I_E$ and $I'_E$. An encryption algorithm is said to posses good diffusion characteristics if there exist a $50\%$ difference in the bits of $I_E$ and $I'_E$.
\subsubsection{Number of Pixel Change Rate(NPCR)}
\label{subsub:npcr}
The NPCR measures the percentage of different pixels in the two images. Higher the value of NPCR, better is the encryption algorithm.
\subsubsection{Unified Average Change in Intensity (UACI)}
\label{subsub:uaci}
It measures the average intensity of differences between the two images. Like NPCR, higher value of UACI indicates better encryption.
\subsection{Miscellaneous Metrics}
\label{subsec:misc}
Members of this category of metrics is used to measure the encryption quality based on a combination of the above mentioned categories.
\subsubsection{Key sensitivity and key space analysis}
\label{subsub:keysp}
One of the significant characteristics of chaotic sequence is having a large key space and high sensitivity to initial conditions. A small change in one or more than one of the values of the input parameters will cause a huge change at the output.  
\subsubsection{Peak Signal to Noise Ration (PSNR)}
\label{subsub:psnr}
This metric measures the robustness of the image encryption algorithm in the presence of noise. PSNR measures the ratio between the maximum possible power of a signal and the power of corrupting noise. Due to the varied dynamic range of different signals, PSNR is usually expressed in terms of the logarithmic decibel scale. For image encryption analysis, the original input image is the signal, and the noise is the error introduced by encryption. A higher PSNR reflects superior encryption quality. 

\subsubsection{Information Entropy}
\label{subsub:ent}
The information entropy is reflects the degree of uncertainties in the system. The information entropy measures the distribution of gray value in the image. The greater information entropy the more uniform of the distribution of gray value.
\subsubsection{Computation Time}
\label{subsub:time}
Apart from the security consideration, running speed of the algorithm is also an important aspect for a good encryption algorithm. The processing time is the time required to encrypt and decrypt an image. The smaller value the processing time has, the better the encryption efficiency will be.

\section{Discussion}
\label{sec:discuss}
We have discussed the recent trends in digital image encryption algorithm based on DNA coding and a chaos theory. The algorithms can be broadly classified into two groups. One group of algorithms have proposed new encryption techniques, and the other cryptanalyze the existing ones. The discussion in Sec. \ref{sub:crynew} reveals various image encryption approaches proposed by the researchers. We have observed that there is a significant amount of overlap among these algorithms. The primary difference among these algorithms is in the use of chaotic functions to generate the pseudo random sequence. For DNA encoding, some techniques uses a predefined binary to DNA mapping, whereas some techniques use dynamic mapping to solve the same problem and thereby introduces more randomness to their algorithm. To show the robustness of their algorithm, authors have used a few or all the evaluation metrics discussed in Sec. \ref{sec:metric}.

Since, there exist no standard datasets to evaluate the image encryption qualities authors have used samples, which are commonly used for other image processing tasks. Among the different algorithms discussed in Sec. \ref{sub:crynew}, only Som et al. \cite{som2013colour} have performed their experiments on two different datasets \cite{weberusc,larson2010most}, while the algorithms in \cite{enayatifar2014chaos,soni2012novel} have experimented with only one dataset \cite{weberusc}. All the other authors have reported their results only on one or two images. This creates a great scope for future researchers to cryptanalyze these algorithms on a variety of other images. Moreover, creation of a standard image dataset to evaluate the encryption qualities of an encryption algorithm also forms a nice scope in the future. 

The works of cryptanalysis evaluate the existing encryption algorithms to reveal the drawback in them. Primarily these cyptanalysis algorithms analyze the diffusion function of a proposed encryption scheme. One approach is to check whether the algorithm is fully reversible or not, i.e. during decryption
is it possible to obtain the original input image. Another approach is to check the resistance of an encryption algorithm in the presence of different statistical attacks, known plain image and chosen plain image attack. 
\section{Summary and Future Work}
\label{sec:summ}
Nowadays, images are the most popular media for information exchange, both in personal and professional environment. Therefore, design and development of a secure and reliable system for information exchange is becoming a necessity. It has applications in various fields, e.g. (i) multimedia communication, (ii) medical imaging, (iii) military communications and many more. This article gives an overview of this rapidly growing field, with a focus on the major techniques available in the literature based on the combination of DNA computing and chaos theory. To the best of our knowledge, ours is the most comprehensive review done on this field, covering the most recent publications. We have also presented the details of the various performance evaluation metrics. Furthermore, the pros and cons of the algorithms are also discussed. This paper gives a good foundation for the researchers currently working in this field as well as the future ones. Reading this paper will give the researchers an appropriate depth as well as the breadth of the topic.

We believe that this is just the beginning of the bio-inspired soft computing research on digital image encryption. In the past researchers have proposed image encryption algorithms based only on DNA computing or chaos theory. But each has its own pros and cons, e.g. chaotic maps have lower key space and are susceptible to be interpreted. This was the rationale behind combining the best practices of these two fields for highly secure and efficient image encryption techniques. Moreover, researchers have applied cellular automata (CA) to various domains, like image encryption, authentication, security. CA is easy to realize in hardware and software. Moreover, it has intrinsic properties suitable for image encryption. Design of a chaos based image encryption using a hybrid model of CA, chaos theory and DNA computing also provides a nice future scope of work. 

In the future, researchers should also look into the feasibility study of implementing these algorithms for practical systems. Since these algorithms are computationally expensive, researchers may better explore the use of parallel processing techniques to reduce the encryption, as well as decryption time. Use of GPU-based clusters on a distributed memory architecture could be a potential solution. In military applications, e.g. satellite images are usually of very large dimension. Instead of linearly processing every pixel, which is highly time-consuming and inefficient, the large image can be divided into blocks and then these individual blocks can be processed on multiple clusters to attain high-performance gain.


%% References


%\bibliographystyle{model1b-num-names}
%\bibliography{surv_bib}

\begin{thebibliography}{00}

%1
\bibitem{el2013image}
El-Samie, Fathi E Abd and Ahmed, Hossam Eldin H and Elashry, Ibrahim F and Shahieen, Mai H and Faragallah, Osama S and El-Rabaie, El-Sayed M and Alshebeili, Saleh A, Image Encryption: A Communication Perspective, CRC Press, (2013)
%2
\bibitem{enayatifar2014chaos}
Enayatifar, Rasul and Abdullah, Abdul Hanan and Isnin, Ismail Fauzi, Chaos-based image encryption using a hybrid genetic algorithm and a DNA sequence, 56, 83--93, (2014)
%3
\bibitem{gao2008new}
Gao, Tiegang and Chen, Zengqiang, A new image encryption algorithm based on hyper-chaos, Physics Letters A, 372, 4, 394--400, (2008)
%4
\bibitem{gehani2004dna}
Gehani, Ashish and LaBean, Thomas and Reif, John, DNA-based cryptography, Aspects of Molecular Computing, 167--188, (2004)
%5
\bibitem{hermassi2014security}
Hermassi, Houcemeddine and Belazi, Akram and Rhouma, Rhouma and Belghith, Safya Mdimegh, Security analysis of an image encryption algorithm based on a DNA addition combining with chaotic maps, 72, 3, 2211--2224, (2014)
%6
\bibitem{jain2013adaptive}
Jain, Anchal and Rajpal, Navin, Adaptive Key Length Based Encryption Algorithm Using DNA Approach, IEEE International Conference on Machine Intelligence and Research Advancement (ICMIRA), (2013)
%7
\bibitem{jain2015robust}
Jain, Anchal and Rajpal, Navin, A robust image encryption algorithm resistant to attacks using DNA and chaotic logistic maps, Multimedia Tools and Applications, 1--18, (2015)
%8
\bibitem{larson2010most}
Larson, Eric C and Chandler, Damon M, Most apparent distortion: full-reference image quality assessment and the role of strategy, Journal of Electronic Imaging, 19, 1, 011006--011006, (2010)
%9
\bibitem{leier2000cryptography}
Leier, Andr{\'e} and Richter, Christoph and Banzhaf, Wolfgang and Rauhe, Hilmar, Cryptography with DNA binary strands, Biosystems, 57, 1, 13--22, (2000)
%10
\bibitem{liu2010color}
Liu, Hongjun and Wang, Xingyuan, Color image encryption based on one-time keys and robust chaotic maps, Computers \& Mathematics with Applications, 59, 10, 3320--3327. (2010)
%11
\bibitem{liu2012image}
Liu, Hongjun and Wang, Xingyuan and others, Image encryption using DNA complementary rule and chaotic maps, Applied Soft Computing, 12, 5, 1457--1466, (2012)
%12
\bibitem{liu2012rgb}
Liu, Lili and Zhang, Qiang and Wei, Xiaopeng, A RGB image encryption algorithm based on DNA encoding and chaos map, Computers \& Electrical Engineering, 38, 5, 1240--1248, (2012)
%13
\bibitem{liu2014cryptanalyzing}
Liu, Yuansheng and Tang, Jie and Xie, Tao, Cryptanalyzing a RGB image encryption algorithm based on DNA encoding and chaos map, Optics \& Laser Technology, 60, 111--115, (2014)
%14
\bibitem{mazloom2009color}
Mazloom, Sahar and Eftekhari-Moghadam, Amir Masud, Color image encryption based on coupled nonlinear chaotic map, Chaos, Solitons \& Fractals, 42, 3, 1745--1754, (2009)
%15
\bibitem{roy2011improved}
Roy, Bibhash and Rakshit, Gautam and Singha, Pratim and Majumder, Atanu and Datta, Debabrata, An improved Symmetric key cryptography with DNA Based strong cipher, IEEE International Conference on Devices and Communications (ICDeCom), (2011)
%16
\bibitem{shannon1949communication}
Shannon, Claude E, Communication theory of secrecy systems*, Bell system technical journal, 28,4,656--715,(1949)
%17
\bibitem{som2013colour}
Som, Sukalyan and Kotal, Atanu and Chatterjee, Ayantika and Dey, Soumista and Palit, Sarbani, A colour image encryption based on DNA coding and chaotic sequences, IEEE International conference on Emerging Trends and Applications in Computer Science (ICETACS), (2013)
%18
\bibitem{soni2012novel}
Soni, Aradhana and Acharya, Anuja Kumar, A novel image encryption approach using an index based chaos and {DNA} encoding and its performance analysis, International Journal of Computer Application, 47, 23, 1--6, (2012)
%19
\bibitem{teng2012bit}
Teng, Lin and Wang, Xingyuan, A bit-level image encryption algorithm based on spatiotemporal chaotic system and self-adaptive, Optics Communications, 285, 20, 4048--4054, (2012)
%20
\bibitem{weberusc}
Weber, Allan G, The USC-SIPI Image Database: Version 5, Signal and Image Processing Institute, University of Southern California, Department of Electrical Engineering, (1997)
%21
\bibitem{wei2012novel}
Wei, Xiaopeng and Guo, Ling and Zhang, Qiang and Zhang, Jianxin and Lian, Shiguo, A novel color image encryption algorithm based on DNA sequence operation and hyper-chaotic system, Journal of Systems and Software, 85, 2, 290--299, (2012)
%22
\bibitem{xie2014breaking}
Xie, Tao and Liu, Yuansheng and Tang, Jie, Breaking a novel image fusion encryption algorithm based on DNA sequence operation and hyper-chaotic system, Optik-International Journal for Light and Electron Optics, 125, 24, 7166--7169, (2014)
%23
\bibitem{xu2012improved}
Xu, Shu-Jiang and Chen, Xiu-Bo and Zhang, Ru and Yang, Yi-Xian and Guo, Yu-Cui, An improved chaotic cryptosystem based on circular bit shift and XOR operations, Physics Letters A, 376, 10, 1003--1010, (2012)
%24
\bibitem{zhang2010image}
Zhang, Qiang and Guo, Ling and Wei, Xiaopeng, Image encryption using  {DNA} addition combining with chaotic maps, Mathematical and Computer Modelling, 52, 11, 2028--2035, (2010)
%25
\bibitem{zhang2013novel_fusion}
Zhang, Qiang and Guo, Ling and Wei, Xiaopeng, A novel image fusion encryption algorithm based on DNA sequence operation and hyper-chaotic system, Optik-International Journal for Light and Electron Optics, 124, 18, 3596--3600, (2013)
%26
\bibitem{zhang2014improved}
Zhang, Qiang and Liu, Lili and Wei, Xiaopeng, Improved algorithm for image encryption based on DNA encoding and multi-chaotic maps, AEU-International Journal of Electronics and Communications, 68, 3, 186--192, 2014
%27
\bibitem{zhang2013novel}
Zhang, Qiang and Wei, Xiaopeng, A novel couple images encryption algorithm based on DNA subsequence operation and chaotic system, Optik-International Journal for Light and Electron Optics, 124, 23, 6276--6281, (2013)
%28
\bibitem{zhang2012novel}
Zhang, Qiang and Xue, Xianglian and Wei, Xiaopeng, A novel image encryption algorithm based on DNA subsequence operation, The Scientific World Journal, 2012, (2012)
%29
\bibitem{zhang2014image}
Zhang, Xuanping and Shao, Liping and Zhao, Zhongmeng and Liang, Zhigang, An image encryption scheme based on constructing large permutation with chaotic sequence, Computers \& Electrical Engineering, 40, 3, 931--941, (2014)
%30
\bibitem{zhou2008parallel}
Zhou, Qing and Wong, Kwok-wo and Liao, Xiaofeng and Xiang, Tao and Hu, Yue, Parallel image encryption algorithm based on discretized chaotic map, Chaos, Solitons \& Fractals, 38,4, 1081--1092, (2008)
%31
\bibitem{zhou2010image}
Zhou, Shihua and Zhang, Qiang and Wei, Xiaopeng, An image encryption algorithm based on DNA self-assembly technology, IEEE International Conference on Intelligent Computing and Intelligent Systems (ICIS), (2010)
%%


\end{thebibliography}



\end{document}